\documentclass{PoS}

\title{Probing Sea Quark and Gluon Polarization at STAR}

\ShortTitle{Probing Sea Quark and Gluon Polarization at STAR}

\author{\speaker{Justin STEVENS, for the STAR Collaboration}\\
       Massachusetts Institute of Technology \\
       E-mail: \email{jrsteven@mit.edu}}

\abstract{ One of the primary goals of the spin program at the Relativistic Heavy Ion Collider (RHIC) is to determine the polarization of the sea quarks and gluons in the proton.  The polarization of the sea quarks is probed through the production of $W^{-(+)}$ bosons via the annihilation of $\bar{u}+d\,(\bar{d}+u)$, at leading order.  In this proceedings we report measurements of the single-spin asymmetry, $A_{L}$, for $W$ boson production at $\sqrt{s} = 510$ GeV, and the new constraints these results place on the antiquark helicity distributions.  Recent results on the longitudinal double-spin asymmetry, $A_{LL}$, for inclusive and di-jet production at $\sqrt{s} = 200$ GeV are also presented.  The inclusive jet results provide the first experimental indication of non-zero gluon polarization in the $x$ range probed at RHIC. }

\FullConference{XXII. International Workshop on Deep-Inelastic Scattering and Related Subjects \\
		 28 April - 2 May 2014 \\
		 Warsaw, Poland}

\newcommand {\MM} [1] {\ensuremath{#1}}

\newcommand {\IT} [1] {\ensuremath{#1}}

\newcommand {\SUB} [2] {\MM{#1\ensuremath{_{#2}}}}
\newcommand {\SUP} [2] {\MM{#1\ensuremath{^{#2}}}}

\newcommand {\momentum} {\IT{p}}

\newcommand {\transverse} {\IT{T}}
\newcommand {\pT} {\SUB{\momentum}{\transverse}}

\newcommand {\program} [1] {\textsc{#1}}

\newcommand {\Wvar} {\IT{W}}
\newcommand {\Wpl} {\SUP{\Wvar}{+}}
\newcommand {\Wmi} {\SUP{\Wvar}{-}}

\newcommand {\RHICBOS} {\program{rhicbos}}
\newcommand {\CHE} {\program{che}}

\begin{document}

\section{Introduction}

Understanding the spin structure of the proton is a fundamental and long standing challenge in nuclear physics.  The contribution of the quark spins to that of the proton has been measured in polarized deep-inelastic scattering (pDIS) experiments, and was found to contribute only $\sim$30\% to the total proton spin \cite{Filippone:2001}.  While the total quark polarization is well determined from inclusive pDIS, flavor separation is accessible in DIS only through semi-inclusive measurements which rely on the use of fragmentation functions to relate measurements of the final-state hadrons to the flavor-separated quark and antiquark distributions.  In addition, the gluon polarization is only indirectly constrained in pDIS through scaling violations.  The antiquark and gluon helicity distributions, extracted from global QCD analyses, therefore have considerably larger uncertainties compared to the well-constrained quark + antiquark sums~\cite{DSSV,LSS}.  It is one of the aims of the STAR spin program to place new stringent constraints on the sea quark and gluon polarizations through the production of $W$ bosons and jets, respectively, in longitudinally polarized proton collisions.

\section{Sea quark polarization}

% W introduction
The flavor-separated helicity-dependent PDFs of the quarks and antiquarks have garnered increased attention due to measurements in the unpolarized sector, most recently by the E866/NuSea Drell-Yan experiment~\cite{E866}.  The E866/NuSea experiment has shown an excess of $\bar{d}$ over $\bar{u}$ quarks in the proton, which has become known as the ``flavor asymmetry'' of the antiquark sea (see Ref.~\cite{Garvey:2001yq} for a review).  This was not anticipated in previous QCD predictions, and brought into question some of the assumptions about the perturbative origin of these sea quarks in the proton.  Several models have been proposed which are consistent with the E866/NuSea data~\cite{Garvey:2001yq}.  Measurements of the flavor-separated antiquark helicity-dependent PDFs at RHIC will provide additional constraints to test these models, and give new insight into the mechanism for generating the light quark sea. 

The $u$ and $d$ quark and antiquark distributions are probed at RHIC through the production of $W^{+(-)}$ bosons, for which the dominant process is the fusion of $u+\bar{d}(d+\bar{u})$ quarks.  The parity-violating nature of the weak production process gives rise to large longitudinal single-spin asymmetries, which provide access to the quark and antiquark helicity distributions.  The asymmetry is defined as $A_L = \Delta\sigma/\sigma$, where $\Delta\sigma = \sigma_+ - \sigma_-$ is the difference in the cross section between positive and negative helicity polarized proton beam, and $\sigma$ is the total cross section.  First measurements of the cross section and spin asymmetry, $A_L$, for $W$ production were reported by the STAR~\cite{PhysRevLett.106.062002,PRDstar} and PHENIX~\cite{PhysRevLett.106.062001} collaborations from data collected in 2009.  The results described in this section (and presented in more detail in Ref.~\cite{Adamczyk:2014xyw}) are from data collected in 2011 (2012) by the STAR experiment with an integrated luminosity of 9 (77)~pb$^{-1}$ at $\sqrt{s}=500$ $(510)$~GeV and an average beam polarization of 0.49 (0.56) with a relative scale uncertainty of 3.4\% on the single beam polarization.

\begin{figure}[th] 
	\includegraphics[width=0.5\textwidth]{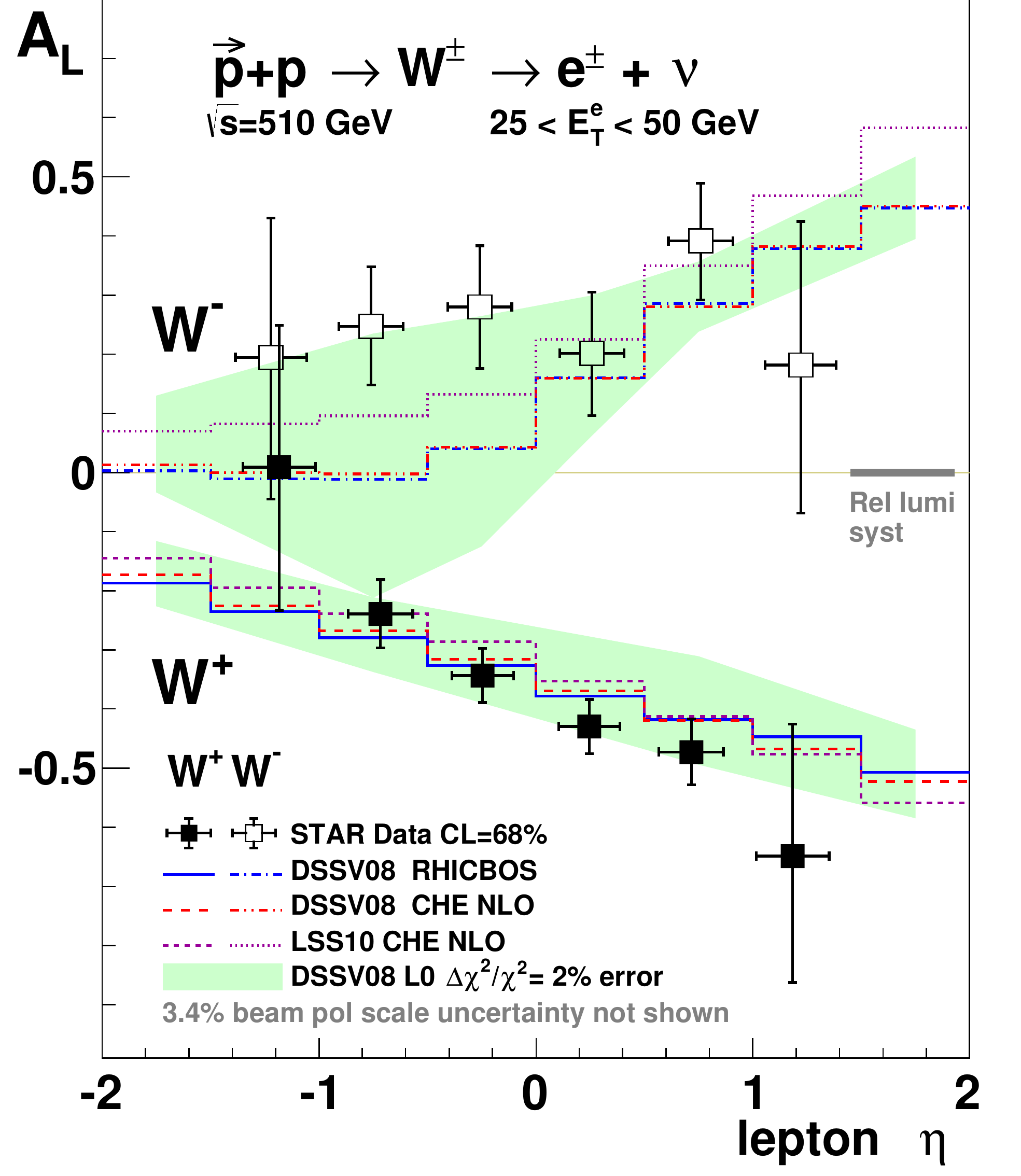} 
	\includegraphics[width=0.45\textwidth]{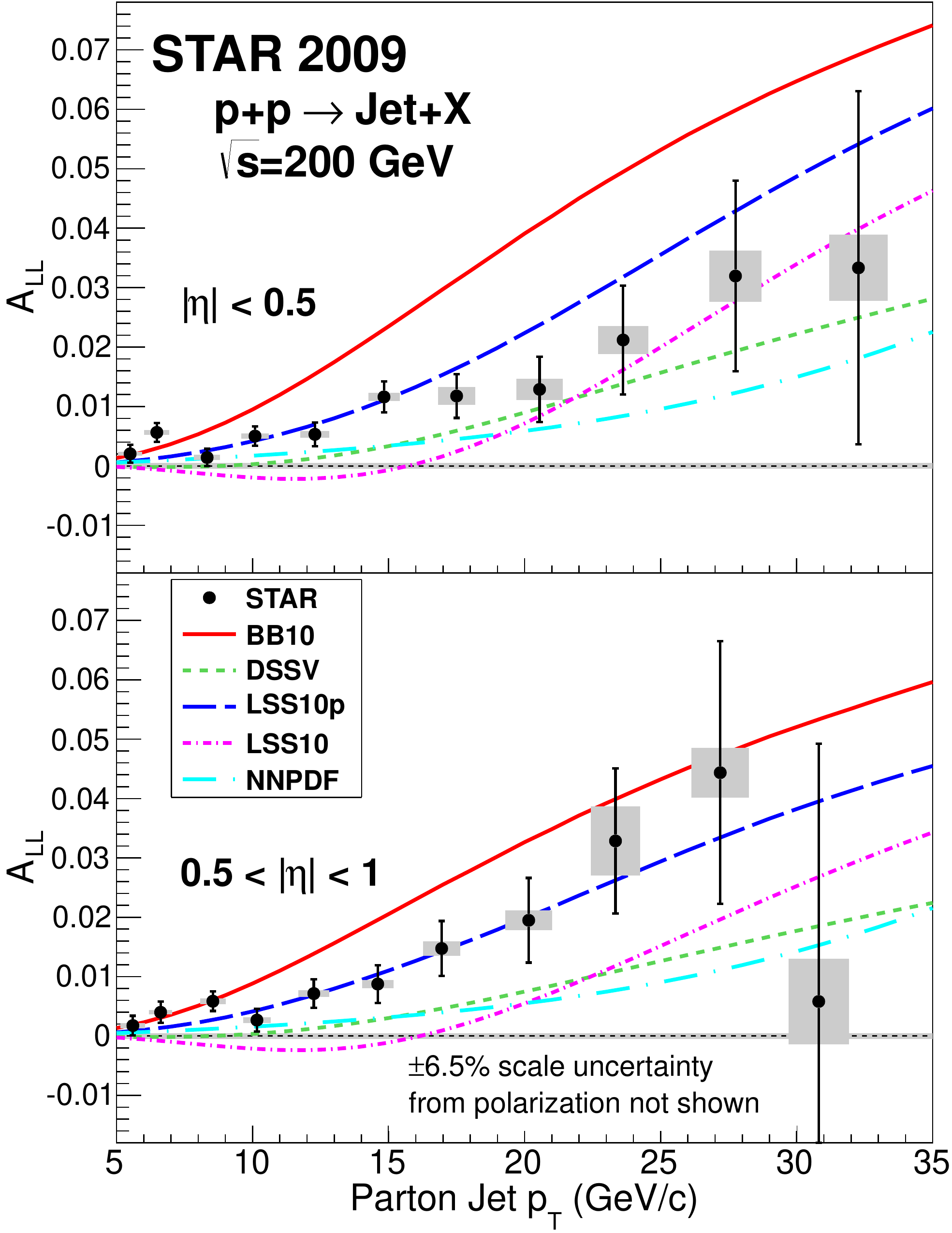}
    	\caption{ (left) Longitudinal single-spin asymmetry, $A_L$, for $W^\pm$ production as a function of lepton pseudorapidity, $\eta_e$.  (right) Inclusive jet $A_{LL}$ as a function of jet \pT{} for midrapidity ($|\eta|<0.5$, upper panel) and forward rapidity ($0.5<|\eta|<1$, lower panel). }
    	\label{fig:jetALL}
\end{figure}

% W results 
As described in Ref.~\cite{Adamczyk:2014xyw}, the spin asymmetries were determined using a profile likelihood method.  The measured single-spin asymmetries are compared to theoretical predictions using both next-to-leading order (\CHE)~\cite{CHE} and fully resummed (\RHICBOS)~\cite{RHICBOS} calculations in Fig.~\ref{fig:jetALL} (left).  The \RHICBOS~calculations are shown for the DSSV08~\cite{DSSV} helicity-dependent PDF set, and the \CHE~calculations are shown for DSSV08~\cite{DSSV} and LSS10~\cite{LSS}.  The DSSV08 uncertainties were determined using a Lagrange multiplier method to map out the $\chi^2$ profile of the global fit~\cite{DSSV}, and the $\Delta\chi^2/\chi^2=2\%$ error band in Fig.~\ref{fig:jetALL} (left) represents the estimated PDF uncertainty~\footnote{DSSV Group, private communication.}.

The measured $A_L^{\Wpl}$ is negative, consistent with the theoretical predictions.  For $A_L^{\Wmi}$, however, the measured asymmetry is larger than the central value of the theoretical predictions for $\eta_{e^-} < 0$.  This region is most sensitive to the up antiquark polarization, $\Delta\bar{u}$, which is not currently well constrained~\cite{DSSV,LSS} as can be seen by the large uncertainty in the theoretical prediction there.  
While consistent within the theoretical uncertainty, the large positive values for $A_L^{\Wmi}$ indicate a preference for a sizable, positive $\Delta\bar{u}$ in the range $0.05~<~x~<~0.2$ relative to the central values of the DSSV08 and LSS10 fits.  Global analyses from both DSSV++~\cite{Aschenauer:2013woa} and NNPDF~\cite{NNPDF} have extracted the antiquark polarizations, using our preliminary measurement from the 2012 dataset.  These analyses quantitatively confirm the enhancement of $\Delta\bar{u}$ and the expected reduction in the uncertainties of the helicity-dependent PDFs compared to previous fits without our data.

\section{Gluon polarization}

% Jet introduction
The measurement of asymmetries directly sensitive to the gluon helicity-dependent PDF was a primary motivation for establishing the spin program at RHIC.  The inclusive jet asymmetry measurement presented in this section is based on data collected in 2009 by the STAR experiment with an integrated luminosity of 20~pb$^{-1}$ at $\sqrt{s}=200$~GeV and an average beam polarization of 0.57 (with a 6.5\% relative uncertainty on the product of the two beam polarizations) representing a nearly 20-fold increase in statistics compared to previous STAR measurements~\cite{Adamczyk:2012qj}.  These results are presented in more detail in Ref.~\cite{Adamczyk:2014ozi}, and only summarized here.  

Jets were reconstructed with the anti-$k_T$ algorithm with a resolution parameter $R=0.6$, using charged particle momenta measured in the Time Projection Chamber and neutral energy deposits in the calorimeters.  The jet $p_T$ reconstructed at this detector level can be corrected to either the particle or parton level.  Detector jets provide contact between the data and simulation while particle jets are formed from the stable final-state particles produced in a collision.  Parton jets are formed from the hard-scattered partons produced in the collision, including those from initial- and final-state radiation, but not those from the underlying event or beam remnants.  We correct the data to the parton jet level because parton jets provide a better representation of the jets in an NLO pQCD calculation. The anti-$k_T$ algorithm with $R$ = 0.6 was used to reconstruct parton jets in simulated PYTHIA events, with the Perugia 0 tune.  The measured asymmetry values are given at the average parton jet $p_T$ for each detector jet $p_T$ bin.  

The asymmetry $A_{LL}$ was evaluated according to
\begin{equation}
A_{LL}=
  \frac{
       \sum \left(P_B P_Y\right) \left( N^{++} - r N^{+-} \right)
  }{
       \sum \left(P_B P_Y\right)^2 \left( N^{++}  + r N^{+-} \right)
  },
\end{equation}
in which $P_{B,Y}$ are the measured beam polarizations, $N^{++}$ and $N^{+-}$ denote the inclusive jet yields for equal and opposite proton beam helicity configurations, and $r$ is the relative luminosity.  Each sum is over individual runs that were 10 to 60 minutes long, a period much shorter than typical time variations in critical quantities such as $P_{B,Y}$ and $r$.  Values of $r$ were measured run-by-run, and range from 0.8 to 1.2.  

The limitation of the STAR trigger to include only neutral calorimetry biases the subprocess fractional contributions ($gg$ vs.\@ $qg$ vs.\@ $qq$) to the inclusive jet yield.  Using PYTHIA to estimate this bias in the subprocesses, the $A_{LL}$ values for detector jets were corrected for trigger and reconstruction bias effects by using the simulation to compare the observed asymmetries at the detector and parton jet levels.

Figure \ref{fig:jetALL} (right) shows the inclusive jet $A_{LL}$ plotted as a function of parton jet $p_T$ for two $\eta$ bins. The vertical size of the shaded uncertainty bands on the $A_{LL}$ points in Fig. \ref{fig:jetALL} reflects the quadrature sum of the systematic uncertainties due to corrections for the trigger and reconstruction bias ($2-55\times{10}^{-4}$) and asymmetries associated with the residual transverse polarizations of the beams ($3-26\times{10}^{-4}$).  Contributions to $A_{LL}$ from non-collision backgrounds were estimated to be less than 2$\%$ of the statistical uncertainty on $A_{LL}$ for all jet $p_T$ bins and deemed negligible.  The relative luminosity uncertainty ($\pm5\times{10}^{-4}$), which is common to all the points, is shown by the gray bands on the horizontal axes.  It was estimated by comparing the relative luminosities calculated with the BBCs and Zero-Degree Calorimeters, and from inspection of a number of asymmetries expected to yield null results. The horizontal size of the shaded error bands reflects the systematic uncertainty on the corrected jet $p_T$.  This includes calorimeter tower gain and efficiency and TPC tracking efficiency and momentum resolution effects.  An additional uncertainty has been added in quadrature to account for the difference between the PYTHIA parton jet and NLO pQCD jet cross sections. The PYTHIA vs.\@ NLO pQCD difference dominates for most bins, making the parton jet $p_T$ uncertainties highly correlated.

% Jet discussion
The theoretical curves in Fig.\@ \ref{fig:jetALL} illustrate the $A_{LL}$ expected for the polarized PDFs associated with the corresponding global analyses. These predictions were made by inserting the polarized PDFs from BB \cite{Blumlein:2010rn}, DSSV \cite{DSSV}, LSS \cite{LSS} and NNPDF \cite{Ball:2013lla} into the NLO jet production code of Mukherjee and Vogelsang \cite{Mukherjee:2012uz}. The BB10 and NNPDF polarized PDFs are based only on inclusive DIS data, while LSS includes both inclusive and semi-inclusive DIS (SIDIS) data sets. LSS provides two distinct solutions for the polarized gluon density of nearly equal quality. The LSS10 gluon density has a node at $x\simeq0.2$, and the LSS10p gluon is positive definite at the input scale $Q^2_0 = 2.5$ GeV$^2$. DSSV is the only fit that incorporates DIS, SIDIS, and previous RHIC $pp$ data.

LSS10p provides a good description of these STAR jet data.  The STAR results lie above the predictions of DSSV and NNPDF and below the predictions of BB10.  However, the measurements fall within the combined data and model uncertainties for these three cases.  In contrast, the STAR jet asymmetries are systematically above the predictions of LSS10 and fall outside the LSS10 uncertainty band for $p_T<15$ GeV/$c$. 

\begin{figure}[h] 
	\begin{centering}
		\includegraphics[angle=-90,origin=c,width=0.483\textwidth]{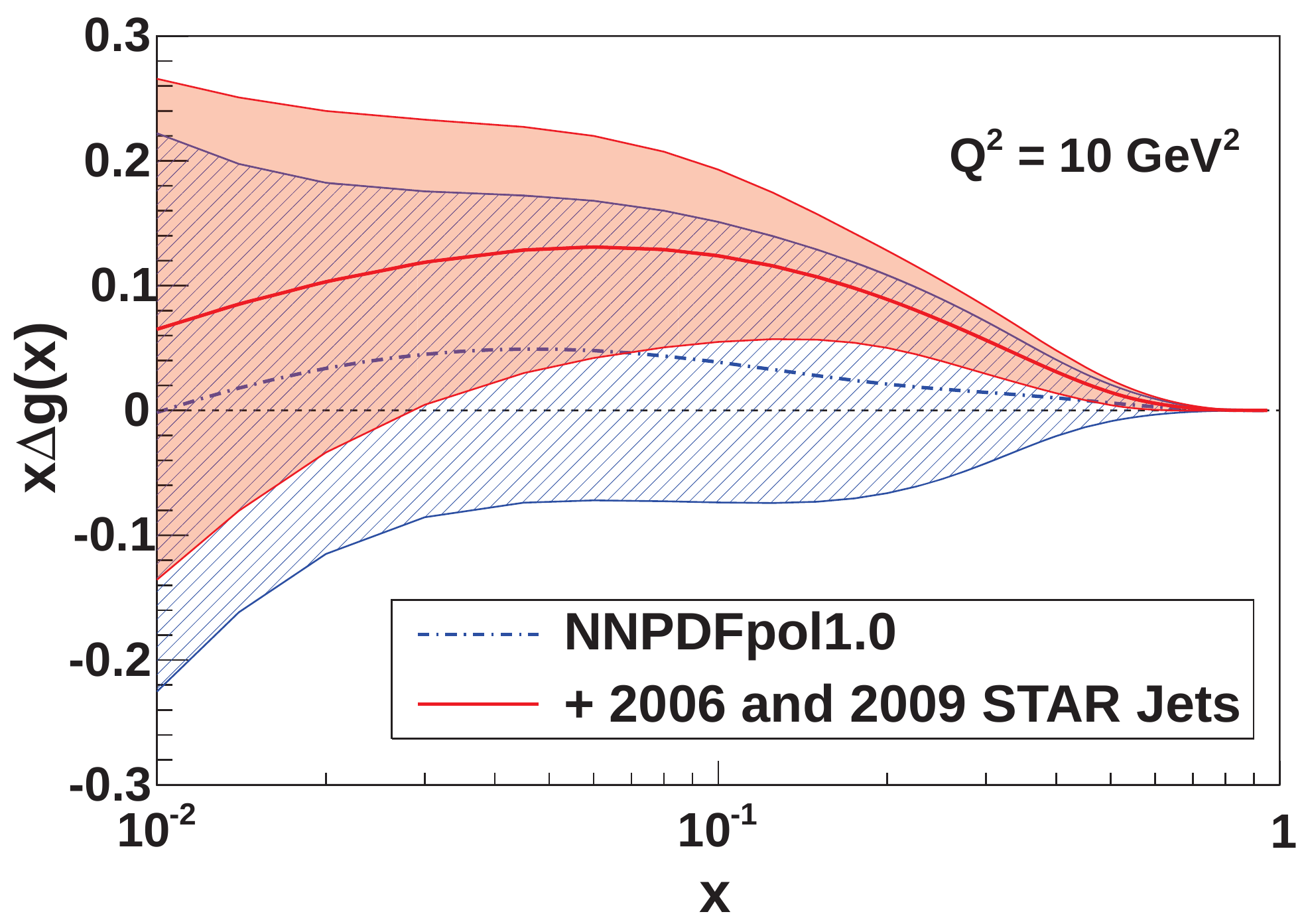} 
		\caption{Uncertainty estimation for the gluon polarization, $\Delta g(x)$, from the Baysian reweighing procedure of the NNPDF group~\cite{Ball:2013lla}. }
    		\label{fig:NNPDF}
	\end{centering}
\end{figure}

The NNPDF group has developed a reweighting method \cite{Ball:2010gb} to include new experimental data into an existing PDF set without the need to repeat the entire fitting process.  The method involves calculating weighted averages over previously equivalent PDF sets, with the weight for each set derived from the $\chi^2$ probability for the set to describe the new data. We have implemented this method to produce a modified NNPDF fit that includes the 2006 \cite{Adamczyk:2012qj} and 2009 STAR jet data.  When calculating the $\chi^2$ probabilities for the jet asymmetries, we included both the statistical and systematic uncertainties and their correlations.  We find that the jet data have a negligible impact on the polarized quark and anti-quark distributions, but a significant impact on the polarized gluon distribution.  Figure \ref{fig:NNPDF} shows the original NNPDF polarized gluon distribution as a function of $x$ at $Q^2$ = 10 GeV$^2$, as well as the modified fit that includes the 2006 and 2009 STAR data.  The integral of $\Delta{g(x,Q^2=10\,\rm{GeV}^2)}$ over the range $0.05<x<0.5$ is $0.06\pm0.18$ for the original NNPDF fit and $0.21\pm0.10$ when the fit is reweighted using the STAR jet data. The inclusion of the STAR jet data results in a substantial reduction in the uncertainty for the gluon polarization in the region $x>0.05$ and indicates a preference for the gluon helicity contribution to be positive in the RHIC kinematic range.

The DSSV group has performed a new global analysis \cite{DSSV:2014} including the STAR jet $A_{LL}$ results presented here.  They find that the integral of $\Delta{g(x,Q^2=10\,\rm{GeV}^2)}$ over the range $x>0.05$ is $0.20^{+0.06}_{-0.07}$ at 90\% C.L., consistent with the value we find by reweighting the NNPDF fit.  DSSV indicates that the STAR jet data lead to the positive gluon polarization in the RHIC kinematic range. The functional form of the polarized parton distribution functions assumed by DSSV is less flexible than that assumed by NNPDF, but DSSV also includes substantially more data in their fit.  Both features may contribute to the smaller uncertainty for DSSV relative to NNPDF.

\section{Conclusion}

In summary, we report new measurements of longitudinal spin asymmetries for $W$ boson and inclusive jet production in polarized $pp$ collisions at $\sqrt{s}$ = 510 and 200 GeV, respectively.  These results provide significant new constraints on both the sea quark and gluon helicity-dependent PDFs when included in updated global analyses.  In particular, the inclusive jet data provide evidence for a positive gluon polarization in the region $x>0.05$.

\end{document}